\documentclass[5p,times,sort&compress]{elsarticle}

\usepackage{hyperref}
\usepackage{siunitx}
\usepackage{cleveref}
\usepackage{xcolor}
\usepackage{nicefrac}
\usepackage{gensymb}
\usepackage{xspace}
\usepackage{tensor}
\usepackage{booktabs}
\usepackage{stfloats}

\newcommand{\GFOUR}{\textsc{Geant4}\xspace}

\makeatletter
\def\ps@pprintTitle{%
   \let\@oddhead\@empty
   \let\@evenhead\@empty
   \def\@oddfoot{\reset@font\hfil\thepage\hfil}
   \let\@evenfoot\@oddfoot}
\makeatother

\begin{document}

\begin{frontmatter}

\title{Extracting low energy signals from raw LArTPC waveforms using deep learning techniques --- A proof of concept}


\author[addrcern]{Lorenzo~Uboldi}
\author[addrniles]{David~Ruth}
\author[addrcmu]{Michael~Andrews}
\author[addrfnal]{Michael~H.~L.~S.~Wang\corref{mycorrespondingauthor}}
\cortext[mycorrespondingauthor]{Corresponding author}
\ead{mwang@fnal.gov}
\author[addrfnal]{Hans-Joachim~Wenzel}
\author[addrfnal]{Wanwei~Wu}
\author[addrfnal]{Tingjun~Yang}
%\author[addrfnal]{Michael~H.~L.~S.~Wang\corref{mycorrespondingauthor}}

\address[addrcern]{CERN, The European Organization for Nuclear Research, 1211 Meyrin, Switzerland}
\address[addrniles]{Niles North High School, Skokie, IL 60077, USA}
\address[addrcmu]{Carnegie Mellon University, Pittsburgh, PA 15213, USA}
\address[addrfnal]{Fermi National Accelerator Laboratory, Batavia, IL 60510, USA}

\begin{abstract}
We investigate the feasibility of using deep learning techniques, in the form of a one-dimensional convolutional neural network (1D-CNN), for the extraction of signals from the raw waveforms produced by the individual channels of liquid argon time projection chamber (LArTPC) detectors.  A minimal generic LArTPC detector model is developed to generate realistic noise and signal waveforms used to train and test the 1D-CNN, and evaluate its performance on low-level signals.  We demonstrate that our approach overcomes the inherent shortcomings of traditional cut-based methods by extending sensitivity to signals with ADC values below their imposed thresholds. This approach exhibits great promise in enhancing the capabilities of future generation neutrino experiments like DUNE to carry out their low-energy neutrino physics programs.
\end{abstract}


\end{frontmatter}

%\linenumbers

\section{Introduction}

The liquid argon time projection chamber (LArTPC) has been successfully deployed in a number of recent and currently running neutrino experiments and is the technology of choice for massive, next-generation neutrino experiments like the Deep Underground Neutrino Experiment (DUNE)~\cite{ref:dunetdrv1}.  Born out of combining the novel three-dimensional imaging capabilities of the time projection chamber (TPC)~\cite{ref:tpc} with the unique properties of liquefied noble gases like liquid argon (LAr), it represents the modern, electronic equivalent of the bubble chamber~\cite{ref:rubbia}.

Unlike the latter, LArTPC detectors are ``always-on" devices, continuously detecting and acquiring signals induced by ionization charges on wire planes at the end of the drift path.  Furthermore, the electronic readout of these signals from multiple wire planes with different angular orientations that provide 2 spatial coordinates, together with a third determined from drift times, enables the automated reconstruction of detailed topologies, while simultaneously performing calorimetry from the integrated charge.

Alongside the unique benefits provided by the TPC are the excellent characteristics of LAr which include high ionization and scintillation yields. The ability to transport electrons efficiently across long drift distances due to the vanishing electronegativity and low dispersion in LAr permits the high spatial resolution possible with finely segmented wire planes. This allows very large detectors like the DUNE far detectors~\cite{ref:dunetdrv1} to be built. Finally, the high density of LAr makes it an ideal neutrino detector due to the low neutrino-nucleon interaction cross sections.

The powerful capabilities of the LArTPC make it an excellent choice for DUNE's long baseline physics program with goals that include determining the neutrino mass hierarchy, observing CP violation in the lepton sector, and making precise measurements of the oscillation parameters using the wide-band beam from Fermilab~\cite{ref:masshrchy,ref:cpvio}.  Performing these measurements with a LArTPC will not be too challenging due to the relatively high incident neutrino energies, ranging from hundreds of MeVs to a few GeVs, and the reduced background levels from cosmogenic and atmospheric sources made possible by the deep underground location of the far detectors.

Beyond its long-baseline program, DUNE's physics goals also include the detection of neutrinos from core-collapse supernovas, searches for nucleon decay, studies of solar neutrinos, and atmospheric neutrino oscillation studies to supplement the long-baseline measurements~\cite{ref:dunesnb,ref:duneprot,ref:dunesolar}.  Of these, the solar and core-collapse supernova neutrinos involve low energy neutrinos in the ~1 MeV (solar) to ~10 MeV (supernova) range.  Ionization from the products of their interactions in the LAr can induce signals that are close to the noise threshold, making them challenging to detect.  This is further exacerbated by the conventional approach of applying minimum ADC threshold cuts to discriminate signal waveforms from noise which results in poor low-energy efficiency.

\begin{figure*}[htbp]
\centering
  \begin{tabular}[t]{cccc}
    \includegraphics[width=0.225\textwidth]{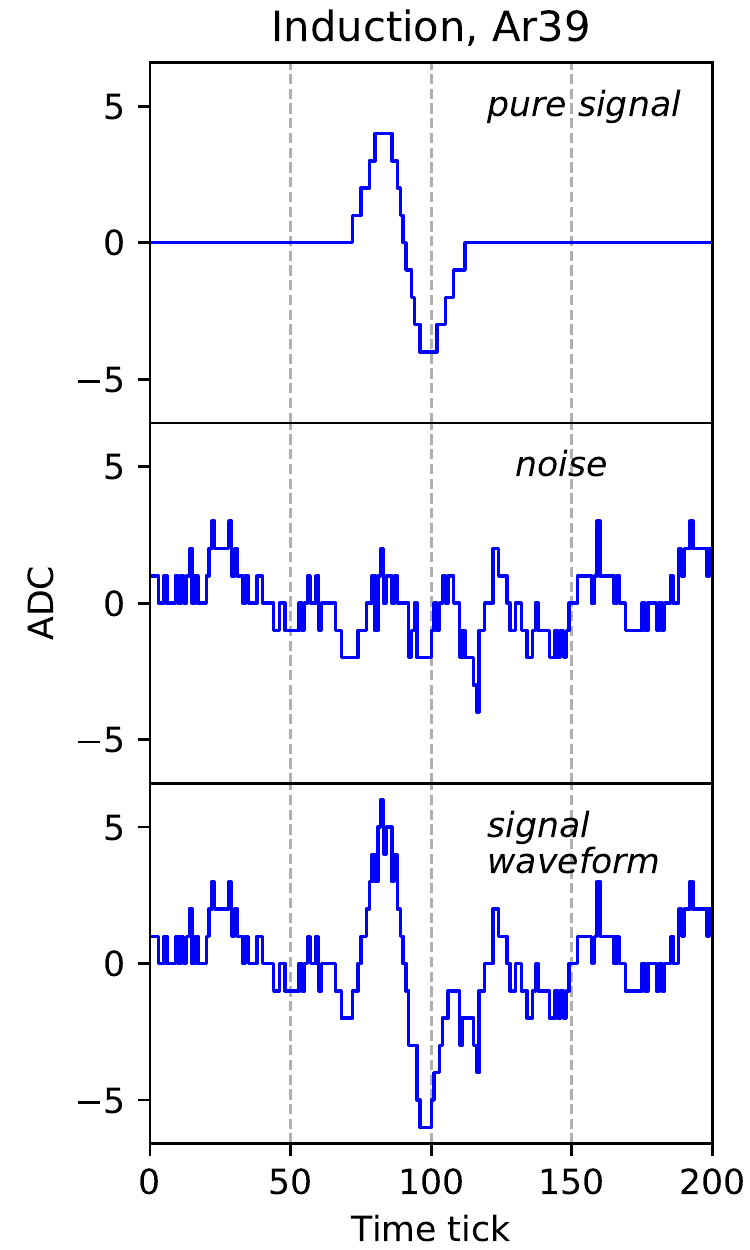} &
    \includegraphics[width=0.225\textwidth]{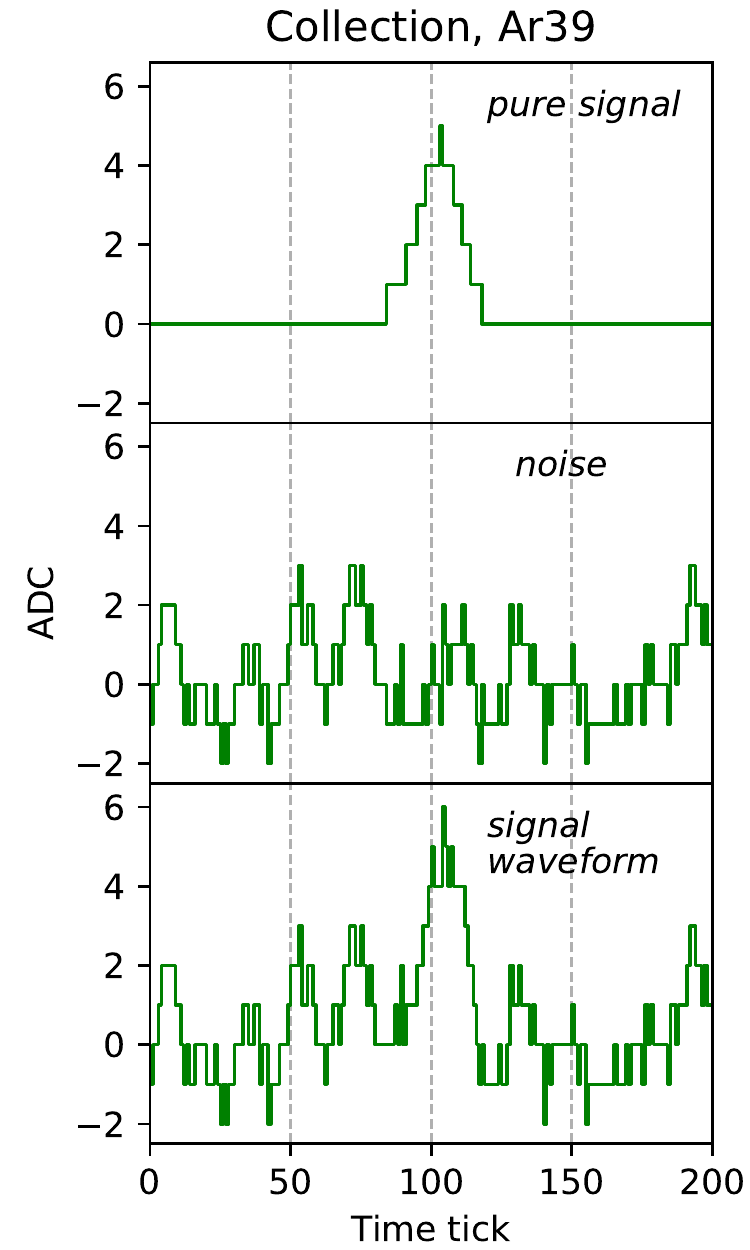} &
    \includegraphics[width=0.225\textwidth]{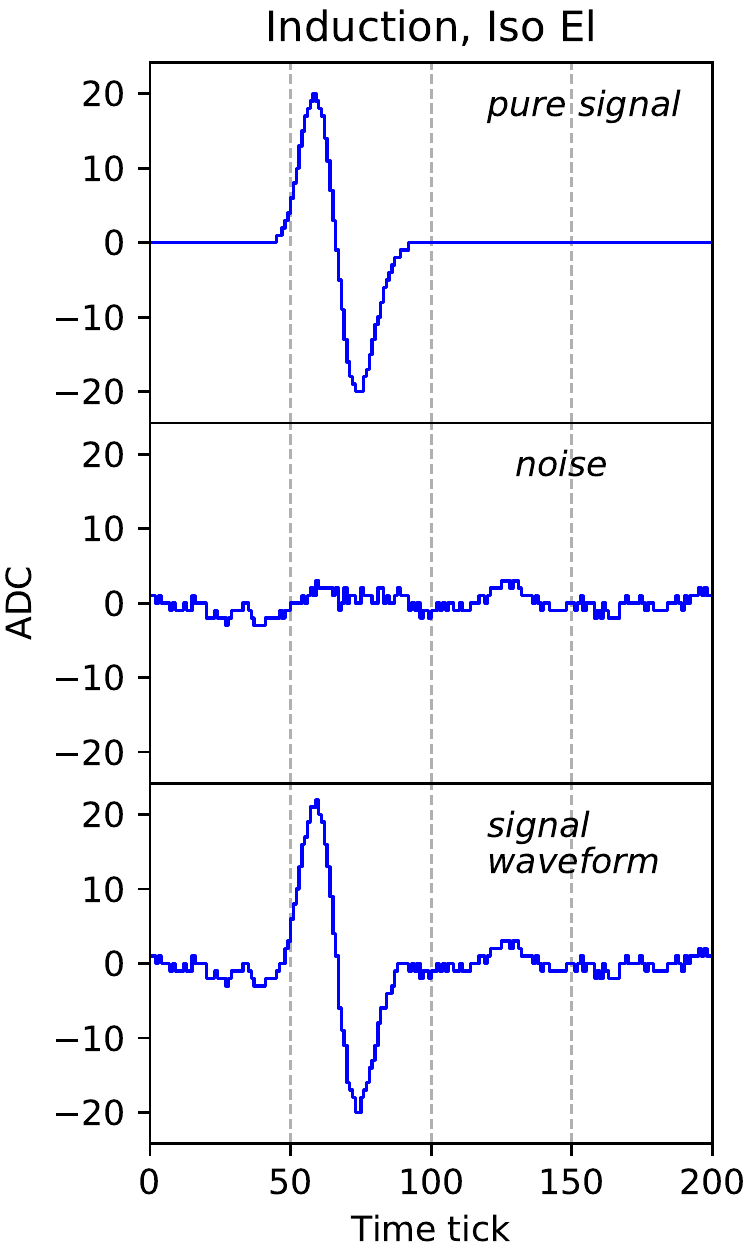} &
    \includegraphics[width=0.225\textwidth]{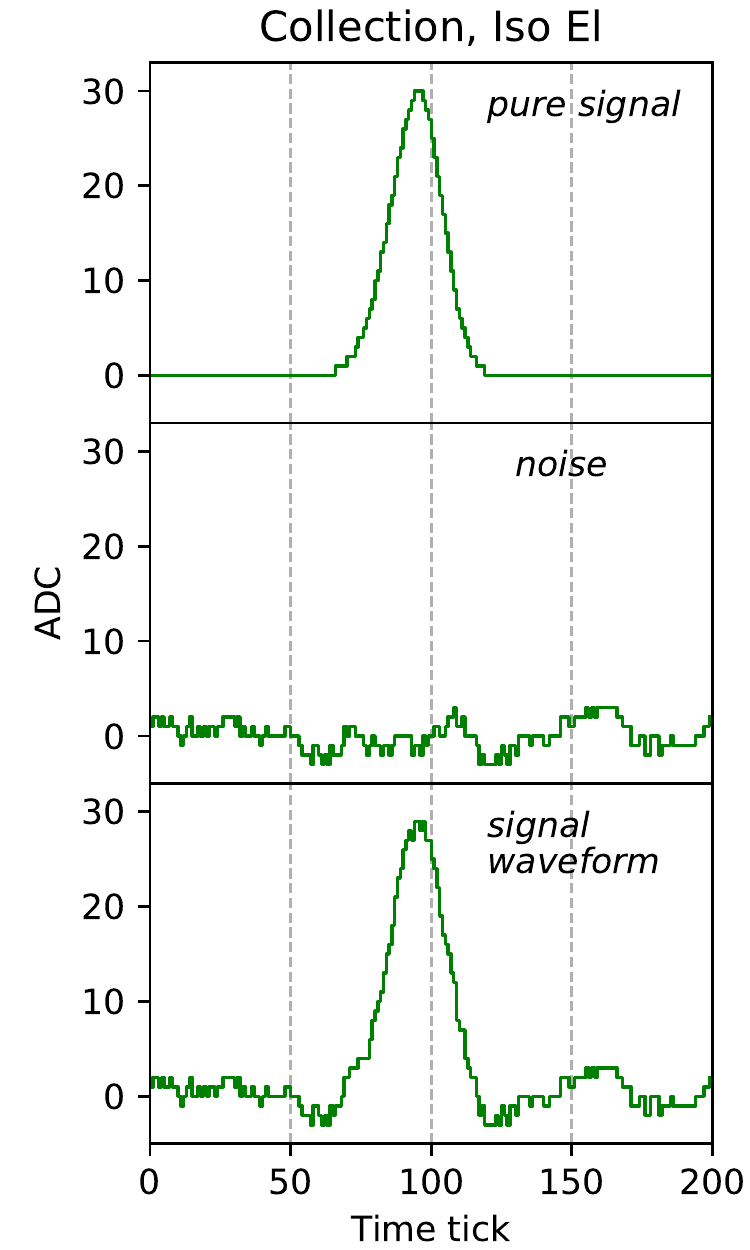} \\
    (a) & (b) & (c)  & (d)\\
  \end{tabular}
\caption{Plots (a) and (b) show simulated wire waveforms for the $\tensor[^{39}]{\mathrm{Ar}}{}$ sample in the induction and collection planes, respectively. Plots (c) and (d) show simulated wire waveforms for the isotropic electron sample in the induction and collection planes, respectively. The first row in each plot shows the pure signal waveform, followed by the noise waveform in the second row, and the sum of these two components in the third row. Details on the simulation of these waveforms is described in Section~\ref{sec:zendet} of the text.}
\label{fig:wavesim}
\end{figure*}

In this paper, we take a deep learning (DL) approach to address the drawbacks of conventional threshold-based methods and to optimize the efficiency to low-energy neutrinos. We develop deep learning techniques and apply them to the raw waveforms from individual LArTPC wires, to detect the presence of a signal and narrow down its location in the full waveform in terms of a region of interest (ROI).  While DL methods have been applied to LArTPC data, they tend to be applied at later stages in offline reconstruction and on 2D ``images" based on wire plane views.  To our knowledge, this is the first attempt to apply such methods directly to the raw waveforms associated with single LArTPC wires.  This implies potentially promising applications of this method in low-level filtering and triggering in online DAQ systems.

In the discussion that follows, we begin by describing a minimalist but realistic \emph{toy} LArTPC detector that we created and used to develop and test our deep learning models.  This is followed by a detailed description of the neural network architecture.  The model's performance on training and validation sets is presented.  It is then tested on an independent sample to determine various performance metrics including the detection efficiency as functions of energy deposited and ADC counts produced. The main purpose of this paper is to demonstrate a proof-of-principle for using deep learning networks in extracting signals from raw LArTPC waveforms.  After successfully accomplishing this, we conclude by summarizing the results and discussing future work and application of the method.

\section{A minimal toy LArTPC detector}\label{sec:zendet}

To test the idea of using DL-based methods for extracting signals from raw LArTPC waveforms, we first developed a software simulation of a minimal \emph{toy} LArTPC detector.  It consists of a LAr volume defined by a rectangular prism measuring $50\times50\times180$ cm$^3$, with the long dimension oriented along the $z$ axis and the two shorter ones along the $x$ and $y$ axes, centered at $(x,y)=(25,0)$.  There are  two instrumented anode wire planes lying in the $y-z$ plane consisting of an induction plane at $x=0.075$ cm and a collection plane at $x=-0.075$ cm.  Both planes consist of 280 equally spaced parallel wires with a pitch of 0.25 cm, and whose axes are oriented at $+60$\degree and $-60$\degree with respect to the $y$ axis for the induction and collection planes, respectively.  A uniform electric field is oriented along the $x$ axis with a field strength of 500 V/cm in the main drift region.

The propagation of particles through the LAr volume was simulated using the \GFOUR simulation toolkit~\cite{ref:gfour1,ref:gfour3} with a step limit of 100 microns through the \emph{LArG4} package in the LArSoft framework~\cite{ref:larsoft}.  If the particle's interaction with the medium in a given step led to energy deposition, the amount of this energy and the position of the interaction was recorded in a \emph{SimEnergyDeposit} object.  This information from the \GFOUR simulation was passed on to the next stage which simulated the propagation of electrons to the readout planes and the response of the detector that ultimately produced the digitized raw wire waveforms from the LArTPC electronics.

\begin{figure*}[htbp]
\centering
\includegraphics[width=0.75\textwidth]{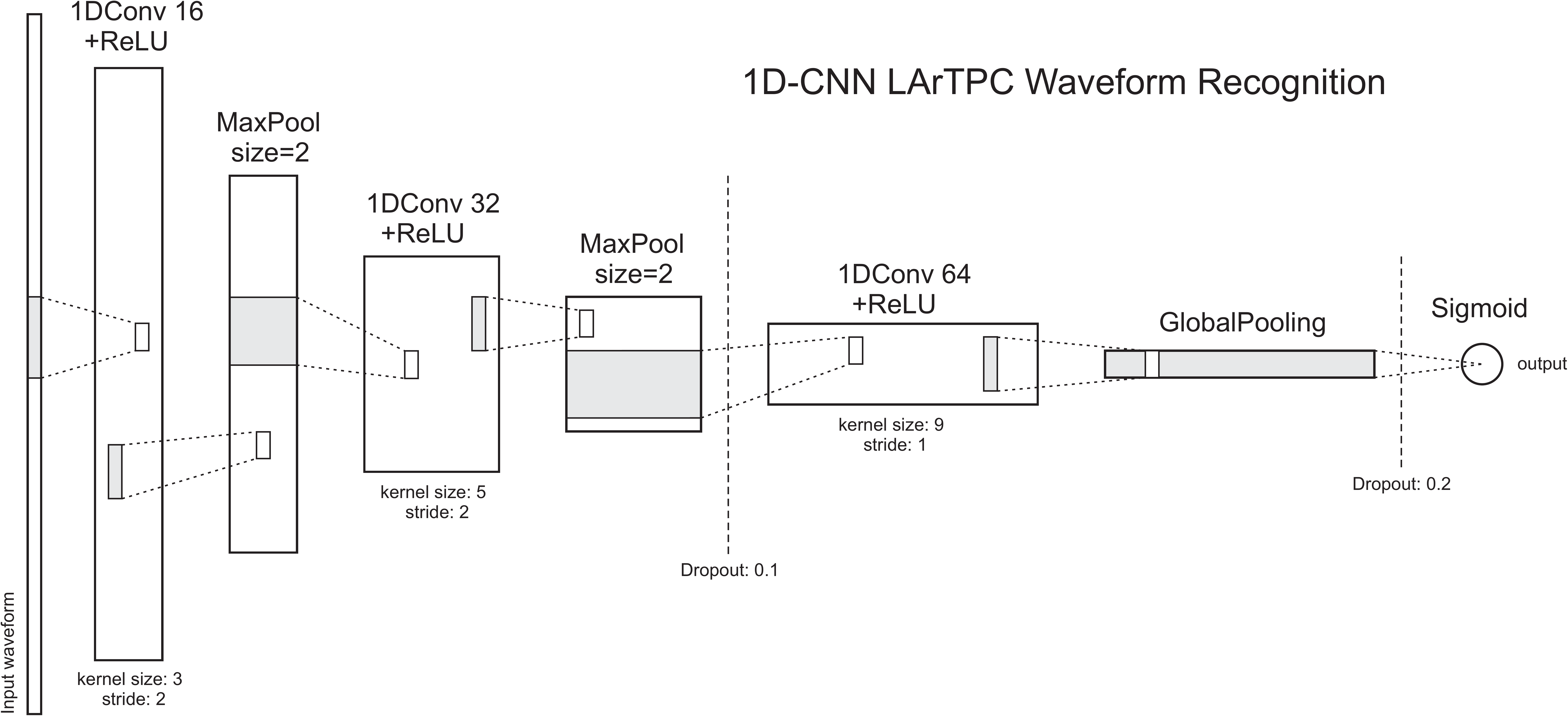}
\caption{Architecture of the one-dimensional convolutional neural network used to recognize signals in LArTPC waveforms.}
\label{fig:1dcnn}
\end{figure*}

The detector response was simulated by first calculating the initial number of ionization electrons resulting from the energy deposition, based on the ionization work function in LAr ($\nicefrac{\mathrm{\#\,electrons}}{\mathrm{GeV}}=\nicefrac{1}{2.36\times10^{-8}\mathrm{GeV}}$)~\cite{ref:wion}.  Electron-ion recombination effects were taken into account by multiplying this initial number by a factor determined from the \emph{box model} of recombination~\cite{ref:rbox} modified to extend into the low $dE/dx$ region~\cite{ref:modrbox}.  Diffusion effects were then simulated by first splitting up the total number of electrons after recombination into clusters of a given size. The initial drift time for all clusters was calculated from the drift velocity and the distance between the original energy deposition and the wire plane. Diffusion in the longitudinal direction was simulated by smearing this drift time according to a Gaussian distribution for each cluster.  Diffusion in the transverse direction was simulated by smearing the transverse positions of the original energy deposition by a Gaussian distribution for each cluster, providing their spatial coordinates at the wire planes.  These smeared coordinates were used to determine the wire closest to a cluster in a plane, on which an induced signal would be simulated. The amount of energy and number of electrons for that cluster were recorded in a \emph{SimChannels} object, and were identified by the channel number corresponding to the wire.  Only channels associated with clusters were stored in the object, and the cluster information was saved in the time bin (TDC) corresponding to its drift time. The waveforms from all LArTPC wires were assumed to be digitized at a rate of 5.05 MHz (198 ns/sample), starting from the instant energy was deposited by the \GFOUR simulated tracks (assumed to all occur simultaneously), and lasting for a duration corresponding to 2,048 samples.

To simulate the waveforms produced by the LArTPC wires, we modeled the field response of the wire planes, as a function of $j$'th TDC bin, with a quadratic ($\propto j^2$) for the collection plane, and with an asymmetric sinusoid, where the amplitude of the negative-going half cycle was 10\% larger, for the induction plane. The electronics response was modeled using the parameterization described in Reference~\cite{ref:eresp}. For each LArTPC wire, the charge in each TDC bin, determined from the total number of ionization electrons stored in the \emph{SimChannels} object for that bin, was convoluted with the field response function for the appropriate wire plane and the electronics response in order to produce what we will refer to as the \emph{pure signal waveform}. Noise was modeled by parameterizing the modulus $r$ of its complex frequency components according to:
\begin{eqnarray}
%p_{0}\times e^{(\frac{x-p_{1}}{p_{2}})^2}
r=|re^{-i\phi}|=p_{0}\times e^{-\frac{1}{2}\left(\frac{f-p_{1}}{p_{2}}\right)^2}\times e^{-\frac{1}{2}\left(\frac{f}{p_{3}}\right)^{p_{4}}}+p_{5}+e^{-p_{6}(f-p_{7})}
\end{eqnarray}
where $f$ represents the midpoint of the $j$'th frequency bin in kHz and the parameters $p_{k}$ are given by $p_{0}=4450$, $p_{1}=-530$, $p_{2}=280$, $p_{3}=110$, $p_{4}=-0.85$, $p_{5}=18$, $p_{6}=0.064$, and $p_{7}=74$.  The modulus $r$ was fluctuated according to a modified Poisson distribution of the form $P(x)=e^{-\mu}\mu^x/(x-1)!$ with $\mu=0.28$, while the phase $\phi$ was generated uniformly from $0\rightarrow2\pi$.  The parameters used in this model were chosen to generate noise very similar to that observed in actual detectors like that described in Reference~\cite{ref:wanweicnnroi}.

The waveform produced by transforming this randomly generated noise into the time domain is what we will refer to as the \emph{noise waveform}, which was generated uniquely for each channel and for every event. The pure signal waveform was added to the noise waveform to produce the simulated LArTPC waveform for a wire if it was associated with a \emph{SimChannels} object; otherwise, only the noise waveform was used. From here on, we will refer to the simulated waveform formed from the sum of a pure signal and a noise waveform simply as a \emph{signal waveform}. Examples of these simulated waveforms are shown in Figure~\ref{fig:wavesim} for the generated samples used in training and testing the DL model described below and in evaluating its performance.

\section{Deep learning approach to LArTPC waveform recognition}

\begin{figure*}[htbp]
\centering
  \begin{tabular}[t]{cccc}
    \includegraphics[width=0.225\textwidth]{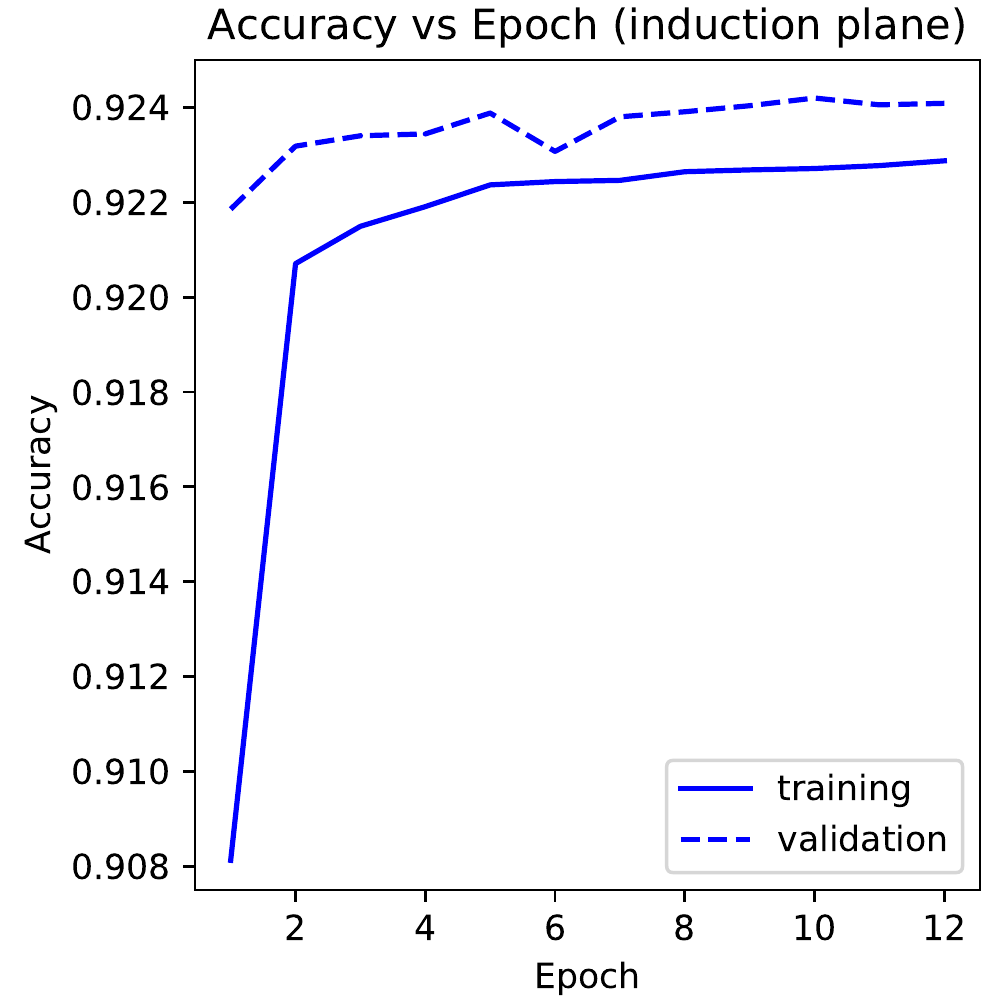} &
    \includegraphics[width=0.225\textwidth]{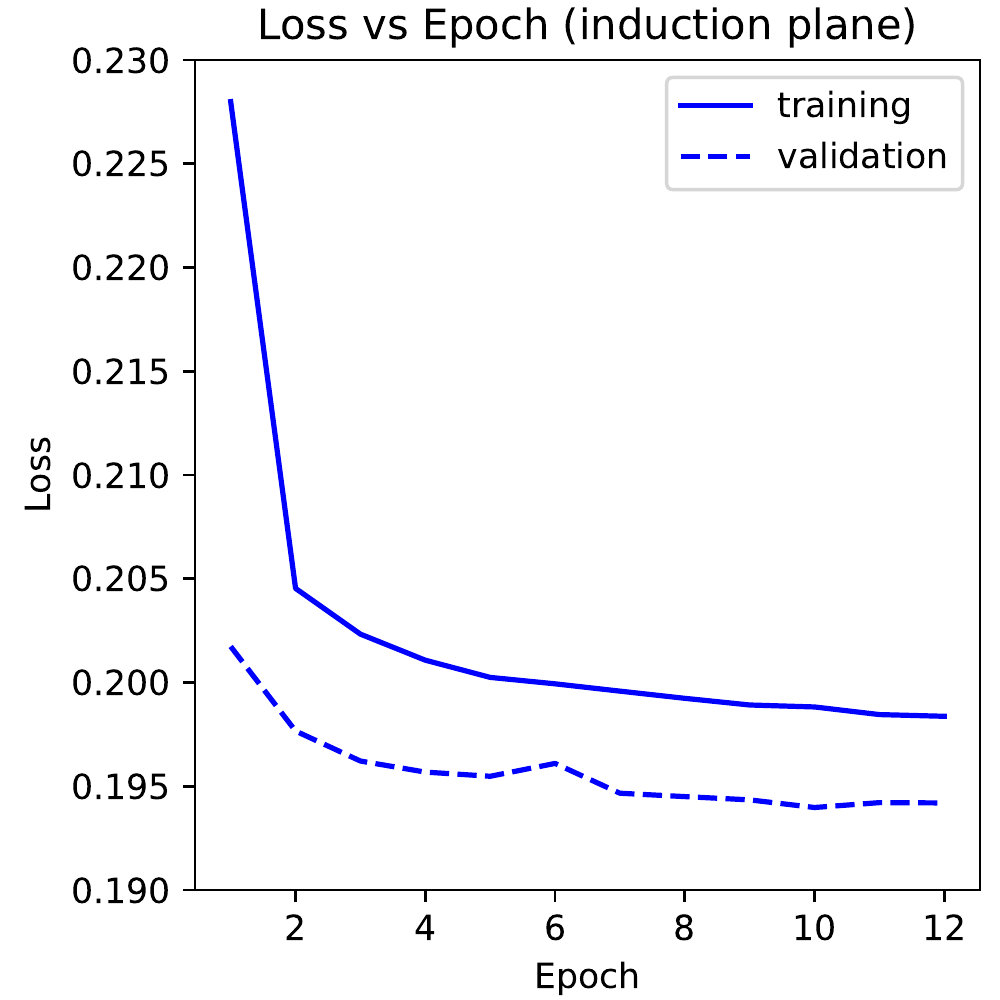} &
    \includegraphics[width=0.225\textwidth]{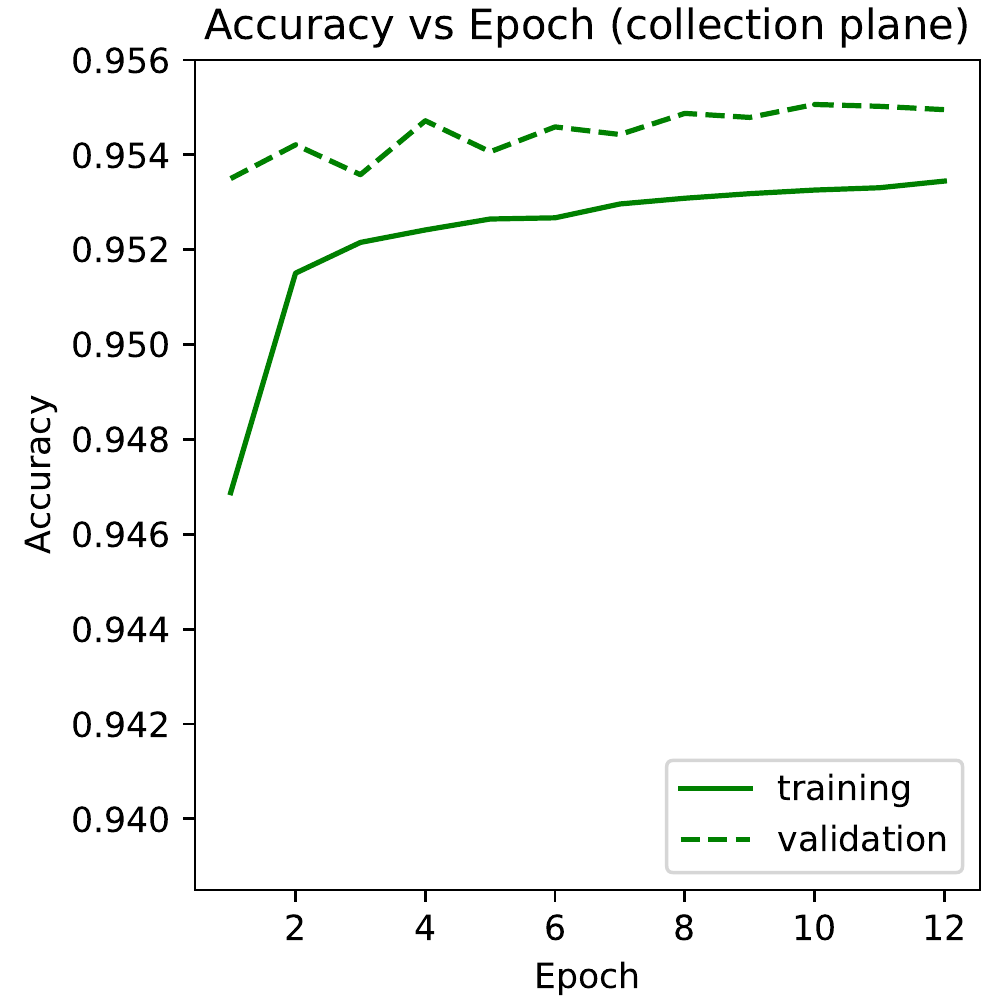} &
    \includegraphics[width=0.225\textwidth]{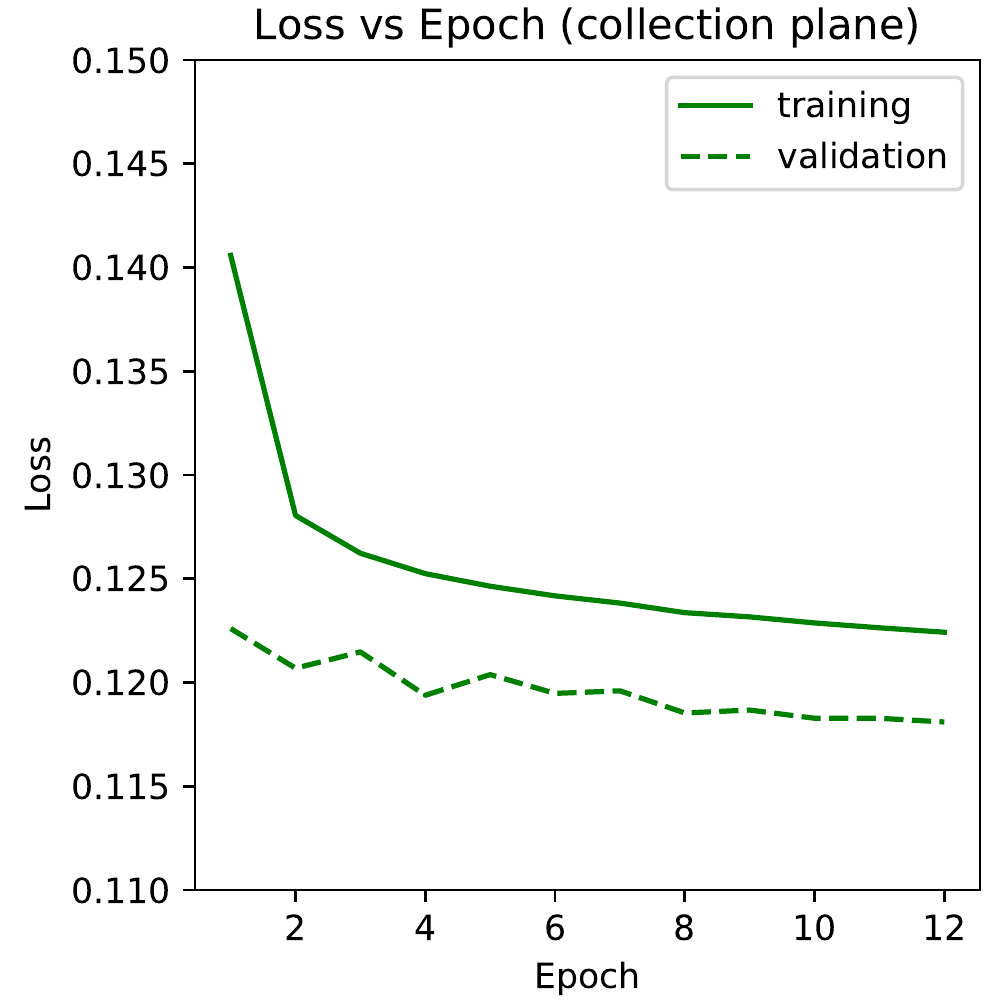} \\
    (a) & (b) & (c)  & (d)\\
  \end{tabular}
\caption{Plots (a) and (b) show how accuracy and loss evolve as a function of training epoch for the induction plane.  Plots (c) and (d) shows the evolution of the same quantities as a function of training epoch for the collection plane.}
\label{fig:acclossvsepoch}
\end{figure*}

\subsection{1D-CNN model architecture}
The architecture of the neural network we developed for detecting signal waveforms in a LArTPC is the one-dimensional convolutional neural network (1D-CNN) shown in Figure~\ref{fig:1dcnn}.  Starting on the left is the waveform presented to the input of the network, which consists of three one-dimensional convolutional layers (Conv1Ds), all of which use rectified linear unit (ReLU) activation functions, defined by $f(x)=\max(0,x)$.  Each Conv1D is immediately followed by a pooling layer that reduces the size of the input feature map.  The first two are maximum pooling layers with pool sizes of 2, and the last is a global pooling layer.  There are 16, 32, and 64 filters or kernels in the first, second, and third Conv1D layers, respectively.  The kernel sizes in this sequence of three layers are 3, 5, and 9, respectively, with stride lengths of 2 for the first two and 1 for the last. Two dropout layers are used, one with a dropout fraction of 0.1 after the second maximum pooling layer, and the other with a dropout fraction of 0.2 after the global pooling layer.  By randomly ignoring a fraction of their inputs, dropout layers help prevent the formation of co-adaptations between layers, which do not generalize well to unseen data and could therefore cause overfitting~\cite{ref:dropout}. The outputs of the global pooling layer terminate into a dense layer with a single node that is activated by a sigmoid function, defined by $f(x)=1/(1+e^{-x})$. This function yields an output bounded between 0 and 1, which can be conveniently interpreted as the probability that the waveform contains a signal or not. This network has a total of 21,217 trainable parameters.

\subsection{Training the model}\label{sec:training}

To create the simulated sample of digitized LArTPC waveforms used to train the model, radiological events from the $\beta$ decay of the $\tensor[^{39}]{\mathrm{Ar}}{}$ nuclide contaminating LAr were first generated. The interactions of the particles produced in these events with the LAr volume were then simulated using the \GFOUR toolkit, followed by the detector response simulation, both of which are described in Section~\ref{sec:zendet}. Next, all the particles in an event that deposited energy in the LAr, leading to a detectable signal from at least one wire, were then identified. This was ensured by requiring the maximum ADC value of the pure signal component of the digitized wire waveform to be $>3$ ADC counts.  In addition, the minimum energy of the parent ionizing particle was required to be $\ge50$ keV, and the maximum number of ionization electrons associated with the signal was required to be $<11,000$.  For each particle identified above, we randomly selected a single wire channel among all those with signals associated with this particle, satisfying the requirements above.  This was done in order to minimize possible correlations between signals from neighboring wires originating from the same particle, which could impact the training process negatively.  Since there can be multiple signal contributions from different particles in a given channel, only the largest contribution, based on the energy deposited, was selected.  A cutout of the full waveform consisting of 200 time bins, with the selected signal region randomly positioned within it, was then created. The training set used in the procedure described below consisted of such 200-tick waveforms.

The model was trained separately for the induction and collection planes of the LArTPC described in Section~\ref{sec:zendet}.  The total number of samples used to train the model was $\approx$2.88 M ($\approx$3.30 M) for the induction (collection) plane.  A separate validation sample of $\approx$721 K ($\approx$824 K) in the induction (collection) plane was not used to train the model directly, but to monitor its performance in the course of training. Both training and validation samples were split roughly equally between signal waveforms and noise waveforms.  Prior to feeding the waveforms to the model, the mean $\bar{x}$ and standard deviation $\sigma$ of all ADC values over all waveforms in the sample were first computed.  Each waveform, identified by the index $i$, was then \emph{standardized} by scaling its ADC values $x_{i}$ by a factor $s_{i}=(x_{i}-\bar{x})/\sigma$.  To fit for the optimal model weights in the training process, we made use of the \emph{Adam} (adaptive moment estimation) optimization algorithm~\cite{ref:adam}. This is an extension of the mini-batch stochastic gradient method that uses per-parameter learning rates, whose values are adapted based on how quickly the weights have been changing. The weights were determined and updated iteratively using random batches of 2,048 waveforms (\emph{batch size}), in which a full pass over the entire sample was completed in one \emph{epoch}.  A total of 8 (12) training epochs were performed for the induction (collection) plane. We used the binary cross-entropy loss function, calculated according to $L=-\frac{1}{N}\sum_{i=1}^Ny_{i}\cdot\log(p_i)+(1-y_i)\cdot\log(1-p_i)$, where the index $i$ runs over the number of observations $N$. $p_i$ is the model output representing the predicted probability for observation $i$ to contain a signal, and $y_i$ is the correct label for that observation (1 for signal and 0 otherwise).

The evolution of the accuracy and the loss as a function of training epoch is shown separately in Figure~\ref{fig:acclossvsepoch} for the induction (blue) and collection (green) planes.  We define $accuracy=\frac{\mathrm{TP}+\mathrm{TN}}{\mathrm{TP}+\mathrm{FN}+\mathrm{TN}+\mathrm{FP}}$, where TP, TN, FP, and FN are the number of true positives, true negatives, false positives, and false negatives, respectively. Loss is as defined above.  For each plane, the accuracy and loss curves are shown for both the training (solid) and validation (dashed) samples.  In all cases, the validation curves follow the general trend of the training curves, reassuring us that overfitting is not an issue.  The values for TN, FP, FN, and TP (elements of the $2\times2$ confusion matrix) at the end of the final training epoch are shown for the training and validation samples in the first two rows of Table~\ref{tab:cm}.

\begin{table*}[htbp]
    \centering
    \begin{tabular}{ccccccccc}
         \toprule
         &\multicolumn{4}{c}{Induction Plane}&\multicolumn{4}{c}{Collection Plane}\\
         \cmidrule(lr){2-5}\cmidrule(lr){6-9}
         Sample & TN & FP & FN & TP & TN & FP & FN & TP \\ \midrule
         training   & 0.477 & 0.023 & 0.052 & 0.448 & 0.483 & 0.017 & 0.028 & 0.472 \\
         validation & 0.476 & 0.024 & 0.052 & 0.448 & 0.482 & 0.017 & 0.028 & 0.473 \\
         testing    & 0.476 & 0.024 & 0.052 & 0.448 & 0.483 & 0.017 & 0.028 & 0.472 \\
         \bottomrule
    \end{tabular}
    \caption{The table above shows the elements of the confusion matrix when applying the fully trained 1D-CNN model on the training, validation, and testing samples. Results are shown for both planes.}
    \label{tab:cm}
\end{table*}

\subsection{Verifying with independent test set}\label{sec:testing}

\begin{figure*}[htbp]
\centering
  \begin{tabular}[t]{cccc}
    \includegraphics[width=0.4\textwidth]{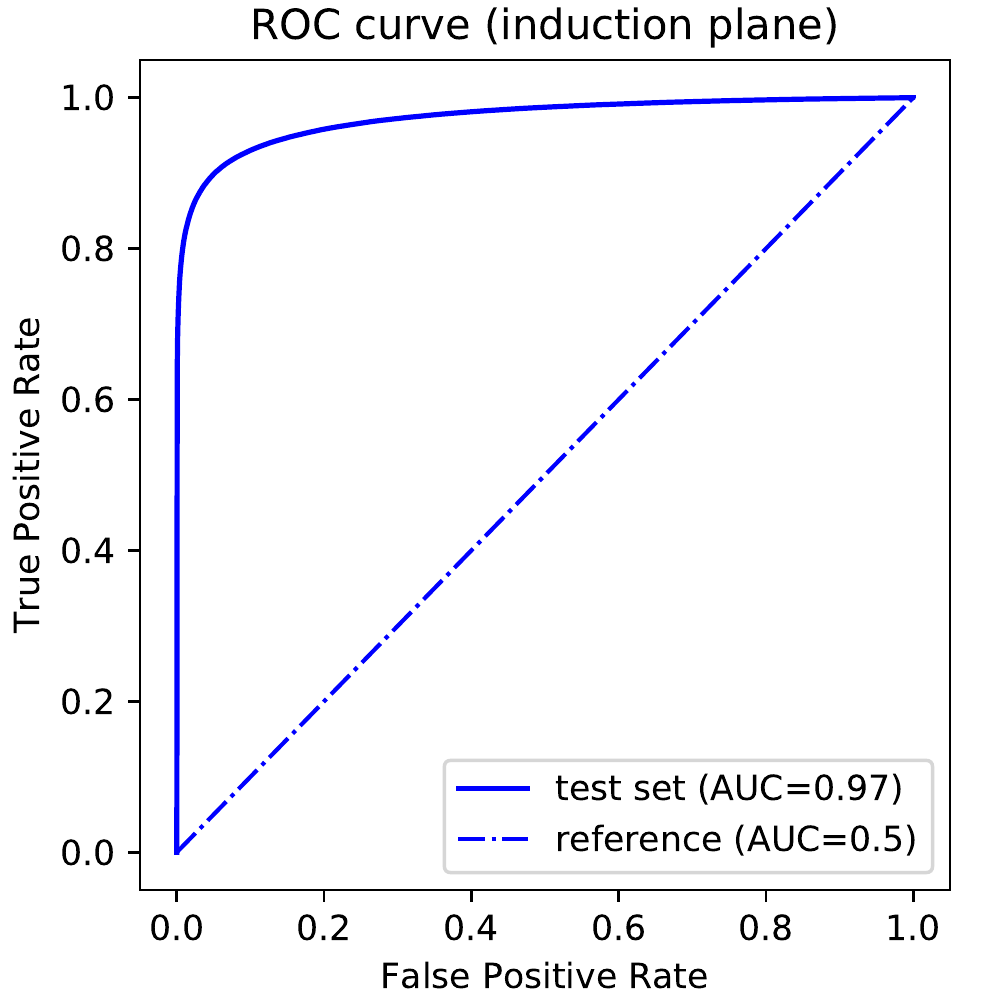} &
    \includegraphics[width=0.4\textwidth]{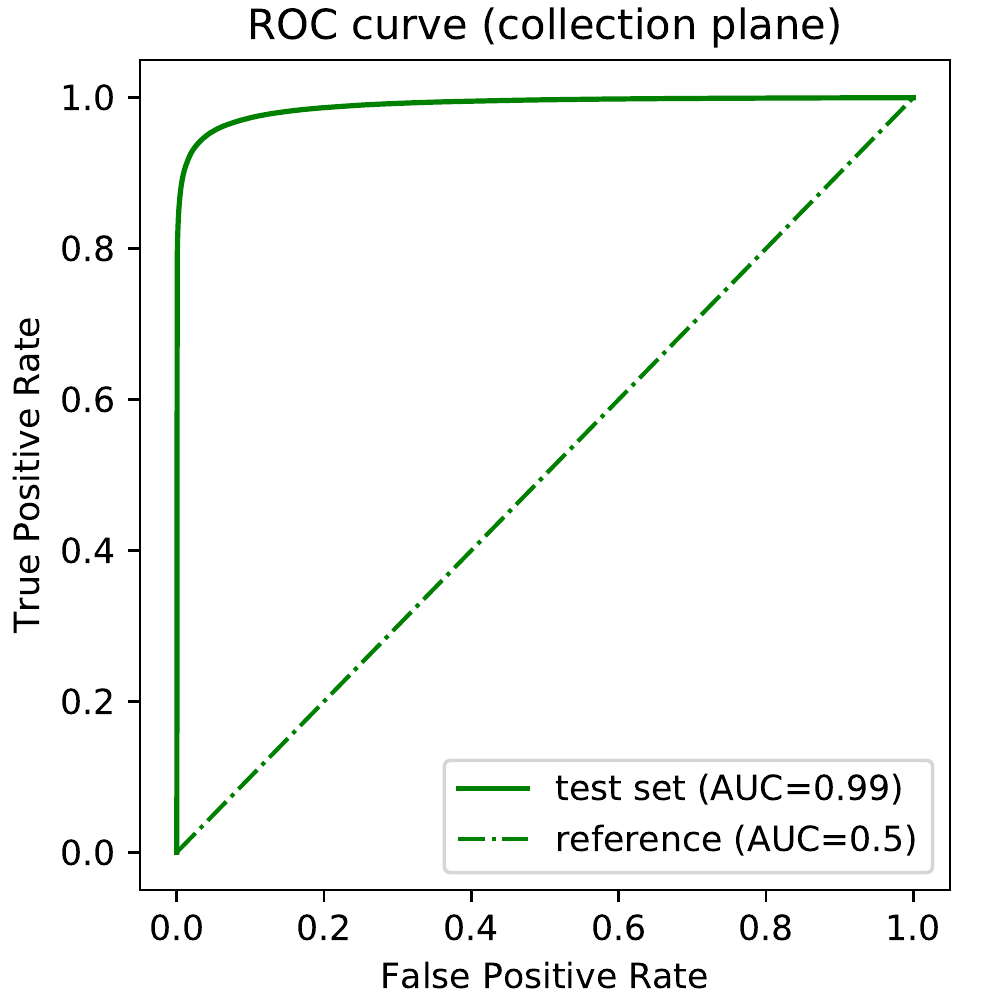} \\
    (a) & (b) \\
  \end{tabular}
\caption{Receiver operating characteristic curves (ROCs) and the associated areas under the ROC curves (AUCs) for the independent test set are shown for the induction plane in (a) and the collection plane in (b).  Shown for reference in each plot is the dash-dotted line representing the case when there is absolutely no discriminating power. }
\label{fig:roccurves}
\end{figure*}

When the training and validation samples were created, a separate and independent test sample of  $\tensor[^{39}]{\mathrm{Ar}}{}$ waveforms was also generated.  This served as an unbiased sample that was not used to train the sample and did not influence the hyperparameter choices for the model.  The size of this sample was $\approx$899 K ($\approx$1.29 M) for the induction (collection) plane, consisting of roughly equal portions of signal and noise waveforms.  We applied the fully trained model described in Section~\ref{sec:training} to the test sample after scaling its ADC values with the same standardization parameters used for the training sample. The results, in terms of the elements of the confusion matrix, are shown in the last row of  Table~\ref{tab:cm}. From these numbers, we calculate 
\emph{precision} $=p=\mathrm{\frac{TP}{TP+FP}}$, \emph{negative predictive value} $=npv=\mathrm{\frac{TN}{TN+FN}}$,
\emph{recall} $=r=\mathrm{\frac{TP}{TP+FN}}$, \emph{specificity} $=s=\mathrm{\frac{TN}{TN+FP}}$, and accuracy as defined previously.  The values for these metrics for the test sample are shown in Table~\ref{tab:metrics}.  In Figure~\ref{fig:roccurves}, the true positive rate is plotted against the false positive rate to show the receiver operating characteristic curves (ROCs) from applying the fully trained model on the test sample for each plane. The diagonal dash-dotted line in each plot represents the reference case when a model has absolutely no ability to tell two classes apart, in which case the area under the ROC curve (AUC) is 0.5. In comparison, our model is able to achieve AUC=0.97 on the induction plane and AUC=0.99 on the collection plane, indicating useful discriminating power between signal and noise waveforms.

\begin{table*}[!bh]
    \centering
    \begin{tabular}{ccc}
    \toprule
    Metric & Induction Plane & Collection Plane \\ \midrule
    precision & 0.950 & 0.965 \\
    negative predictive value & 0.901 & 0.946 \\
    recall & 0.896 & 0.944 \\
    specificity & 0.953 & 0.966 \\
    accuracy & 0.924 & 0.955 \\
    \bottomrule
    \end{tabular}
    \caption{The table above summarizes the performance of the fully trained 1D-CNN model on the test sample.}
    \label{tab:metrics}
\end{table*}

\section{Region of interest finding and model performance}

\subsection{Extending the model to localize signals}\label{sec:roi}

\begin{figure*}[htbp]
\centering
  \begin{tabular}[t]{cccc}
    \includegraphics[width=0.48\textwidth]{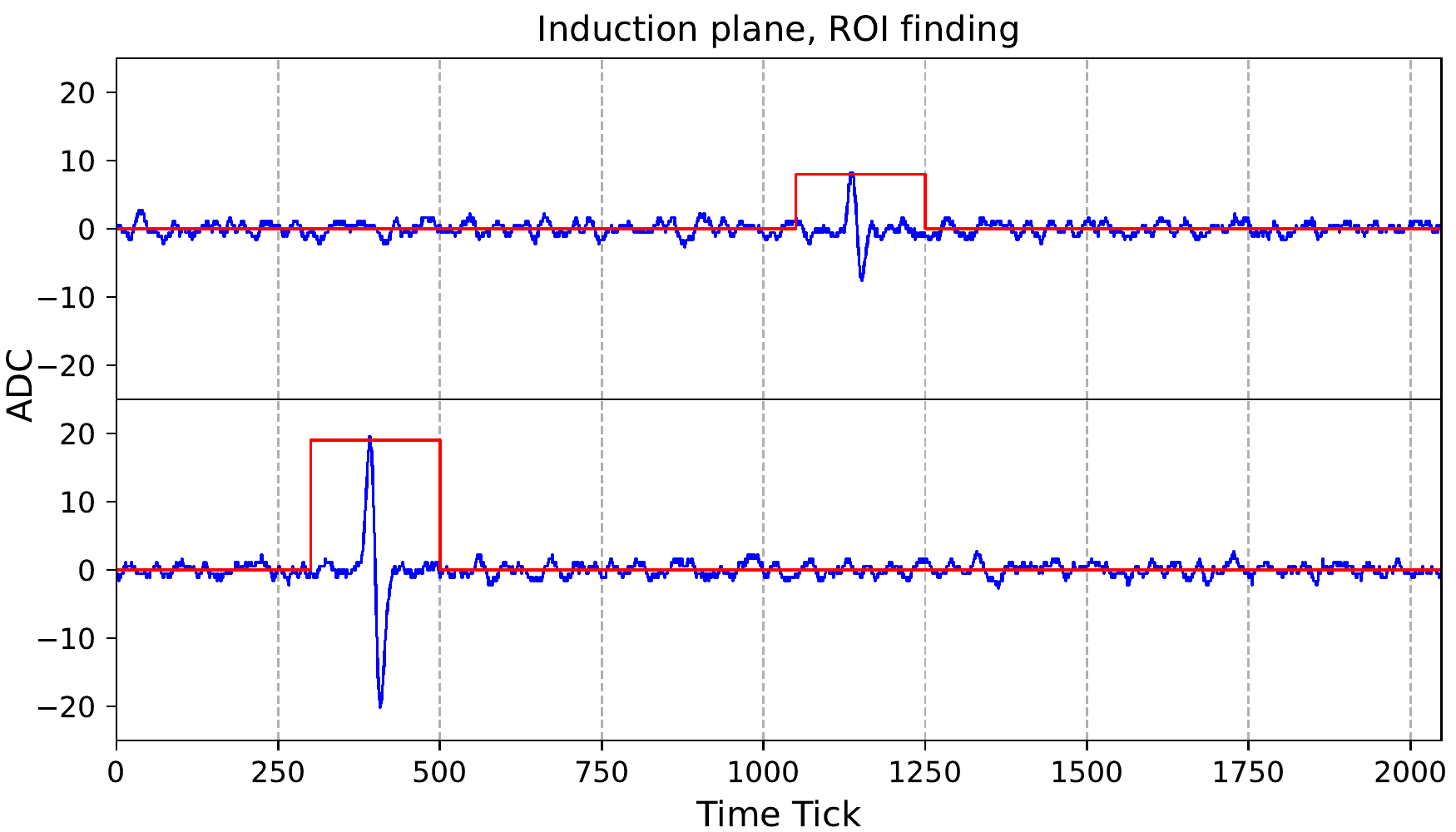} &
    \includegraphics[width=0.48\textwidth]{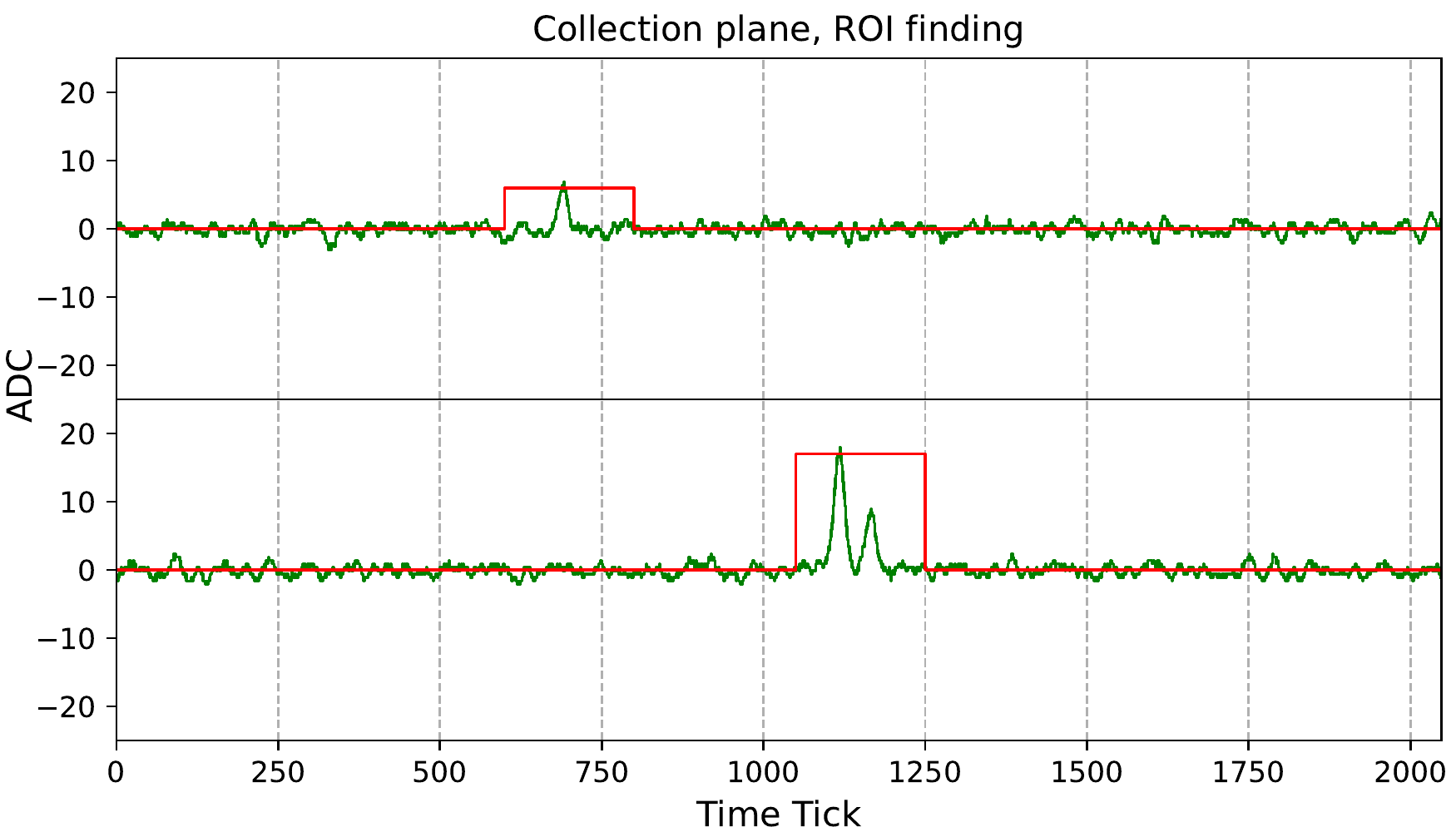} \\
    (a) & (b) \\
  \end{tabular}
\caption{Two examples each of simulated full (2,048-tick) waveforms are shown for the (a) induction and the (b) collection planes.  \emph{Regions-of-interest} (ROIs) that localize the signals within the full waveforms, using the method described in Section~\ref{sec:roi}, are indicated by the rectangular pulses drawn in red. }
\label{fig:roi}
\end{figure*}

The results presented in Section \ref{sec:testing} demonstrate that our 1D-CNN model can make useful predictions about whether 200-tick waveform snippets contain signals or not.  This section describes how we extended this capability to the localization of signals within the full waveforms in terms of ROIs.  This was done in a straightforward manner by simply scanning a window across the entire waveform starting from the leftmost edge and shifting it repeatedly to the right with some finite stride length until it reached the rightmost edge.  By performing an inference on the portion of the waveform contained within the window at each step, the signal region could effectively be localized. In our implementation in this paper, the full 2,048-tick simulated LArTPC waveform was subdivided into 14 overlapping 200-tick windows, beginning with the first, whose left edge was aligned with the start of the waveform.  Each of the 12 subsequent windows after the first was offset from the previous one by a stride length of 150 ticks, while the last window (13th after the first) was offset from the preceding one by 48 ticks, so that its right edge aligned with the last tick of the full waveform.  An inference was then performed on each of these 14 windows to identify the ROI/s within the full waveform. Examples of such ROIs are shown in Figure~\ref{fig:roi}.

\subsection{Evaluating efficiency with single electrons}

\begin{figure*}[htbp]
\centering
  \begin{tabular}[t]{cccc}
    \includegraphics[width=0.49\textwidth]{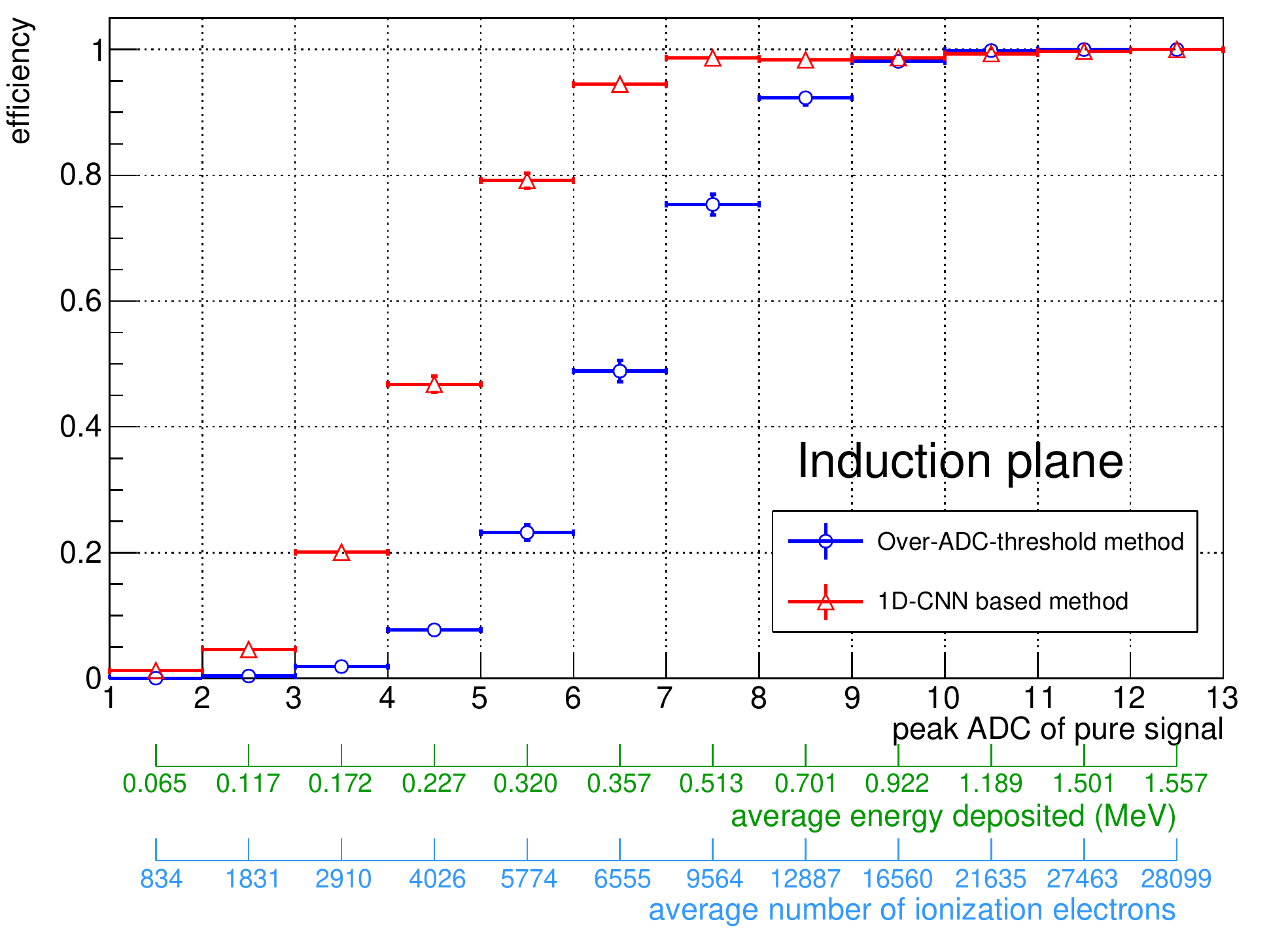} &
    \includegraphics[width=0.49\textwidth]{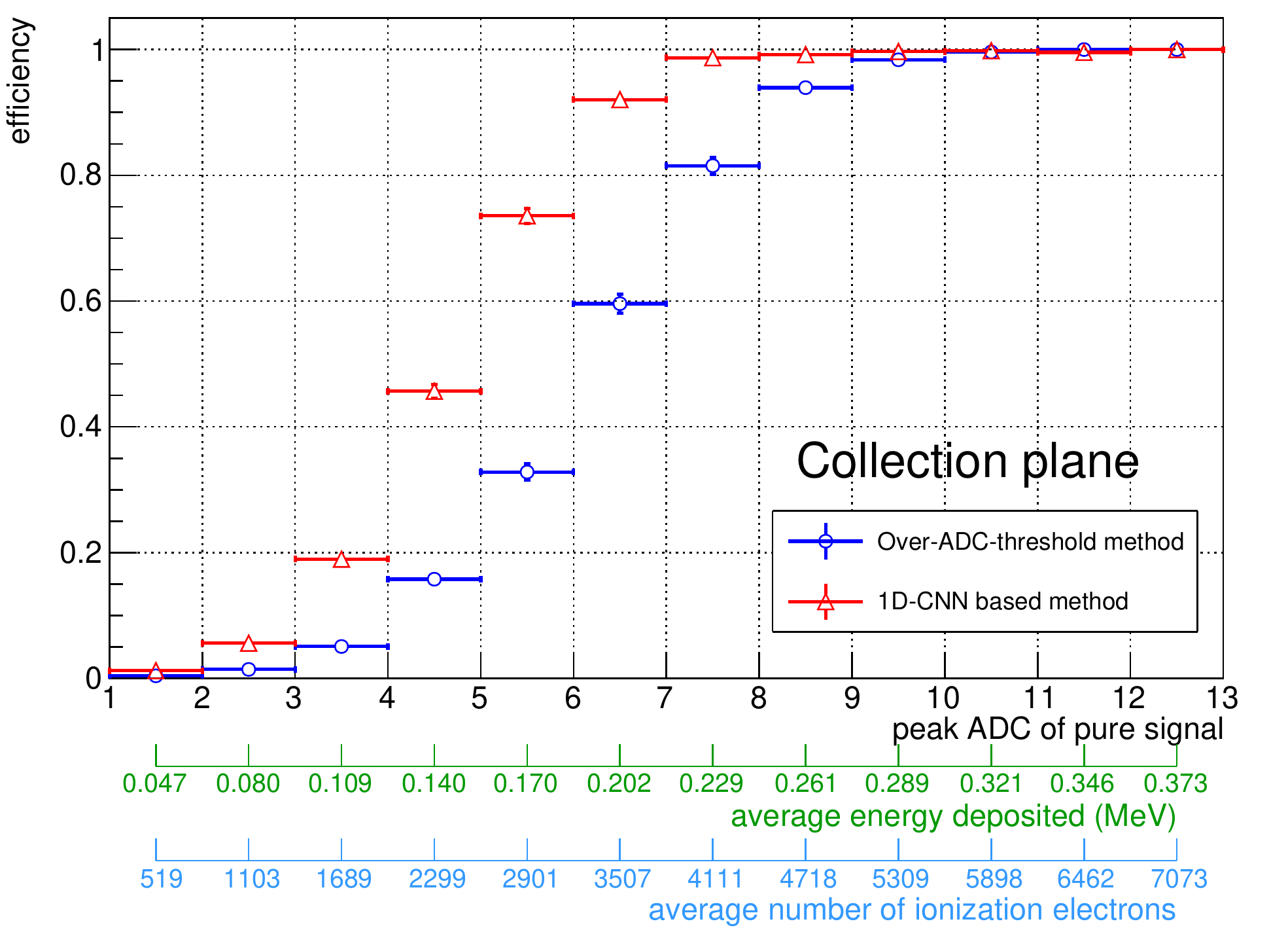} \\
    (a) & (b) \\
  \end{tabular}
\caption{Efficiencies of the 1D-CNN ROI finder as a function of $ADC_{\mathrm{pk}}$ determined using the isotropic electron sample for the (a) induction and (b) collection planes.  Results are shown in each case for the traditional \emph{over-ADC-threshold} and the 1D-CNN based approaches. The average energy deposited in MeV and the average number of electrons associated with the signal, for each $ADC_{\mathrm{pk}}$ bin, are shown below each plot.}
\label{fig:roieff}
\end{figure*}

To further quantify the performance and capabilities of our 1D-CNN model, we generated a single electron sample with electron momenta ranging from 1 MeV to 1 GeV.  The initial position of each electron was generated uniformly within a rectangular prism that centered on the LAr volume and which measures $30\times30\times90$ cm$^3$, with the long dimension oriented along the $z$ axis and the two shorter ones along the $x$ and $y$ axes. The initial direction of each electron was generated pseudo-isotropically with angles $\theta_{xz}$ and $\varphi_{yz}$ distributed uniformly within $\pm180\degree$ and $\pm90\degree$, respectively. After this, the electrons were propagated through the detector using \GFOUR and the detector response simulated as described in Section \ref{sec:zendet}, to produce the raw LArTPC wire waveforms.  The fully trained 1D-CNN, described in Section~\ref{sec:training}, was then applied to each waveform to find the signal ROI as detailed in Section~\ref{sec:roi}.

The signal detection efficiency for the isotropic electron sample is presented here as a function of the peak ADC value associated with the pure signal component of a digitized waveform, which shall be referred to hereon as $ADC_\mathrm{pk}$.  We choose to work with ADC values because they are ultimately what the 1D-CNN directly ``sees''.  This avoids the need to convert from some other quantity like energy to ADC values, which would be experiment specific.

To calculate efficiency, two histograms are created to represent the numerator ($H_\mathrm{N}$) and denominator ($H_\mathrm{D}$) of the ratio. Each histogram has 12 bins representing $ADC_\mathrm{pk}$ ranging from 1 to 12.  If a signal is present within the full 2,048-tick waveform, an entry is made in the $H_\mathrm{D}$ bin corresponding to the $ADC_\mathrm{pk}$ associated with the signal. In case multiple signals are present within the full waveform, only the one with the largest $ADC_\mathrm{pk}$ is selected for calculating efficiency. Following the procedure described in Section~\ref{sec:roi}, the 1D-CNN is used to scan the entire waveform for ROIs.  If the signal lies within an ROI identified by the 1D-CNN, an entry is also made in the same bin of $H_\mathrm{N}$. Once the two histograms are filled, their ratio is taken to yield the efficiency as a function of $ADC_\mathrm{pk}$.  This is done separately for the induction and collection planes, and the results are shown in red in Figure~\ref{fig:roieff}.  The average energy deposited and average number of ionization electrons associated with the signal, for each bin of $ADC_\mathrm{pk}$, are also shown below each plot. The CNN output is required to be $>0.91$ ($>0.95$) in order to classify a waveform as a  signal in the induction (collection) plane.

Also shown for comparison in each plot are the efficiencies (in blue) for the traditional \emph{over-ADC-threshold} approach.  In this case, signal waveforms are discriminated from noise by requiring the measured ADC values in a region containing the signal to be above noise threshold. The mean ADC value of the pure noise waveforms generated as described in Section~\ref{sec:zendet} is 1.77, with a standard deviation of 1.05. In these plots, we require $\left|\mathrm{ADC}\right| > 6$ counts for the induction plane and  $\mathrm{ADC}>6$ counts for the collection plane. These correspond to $\approx 4$ standard deviations above the noise average.  Such a requirement achieves a background rejection rate of 0.926 (0.962) on the induction (collection) plane, which closely matches the rate of 0.931 (0.966) from the CNN-based method when requiring the output to be $>0.91$ ($>0.95$).

\begin{figure*}[htbp]
\centering
  \begin{tabular}[t]{cccc}
    \includegraphics[width=0.49\textwidth]{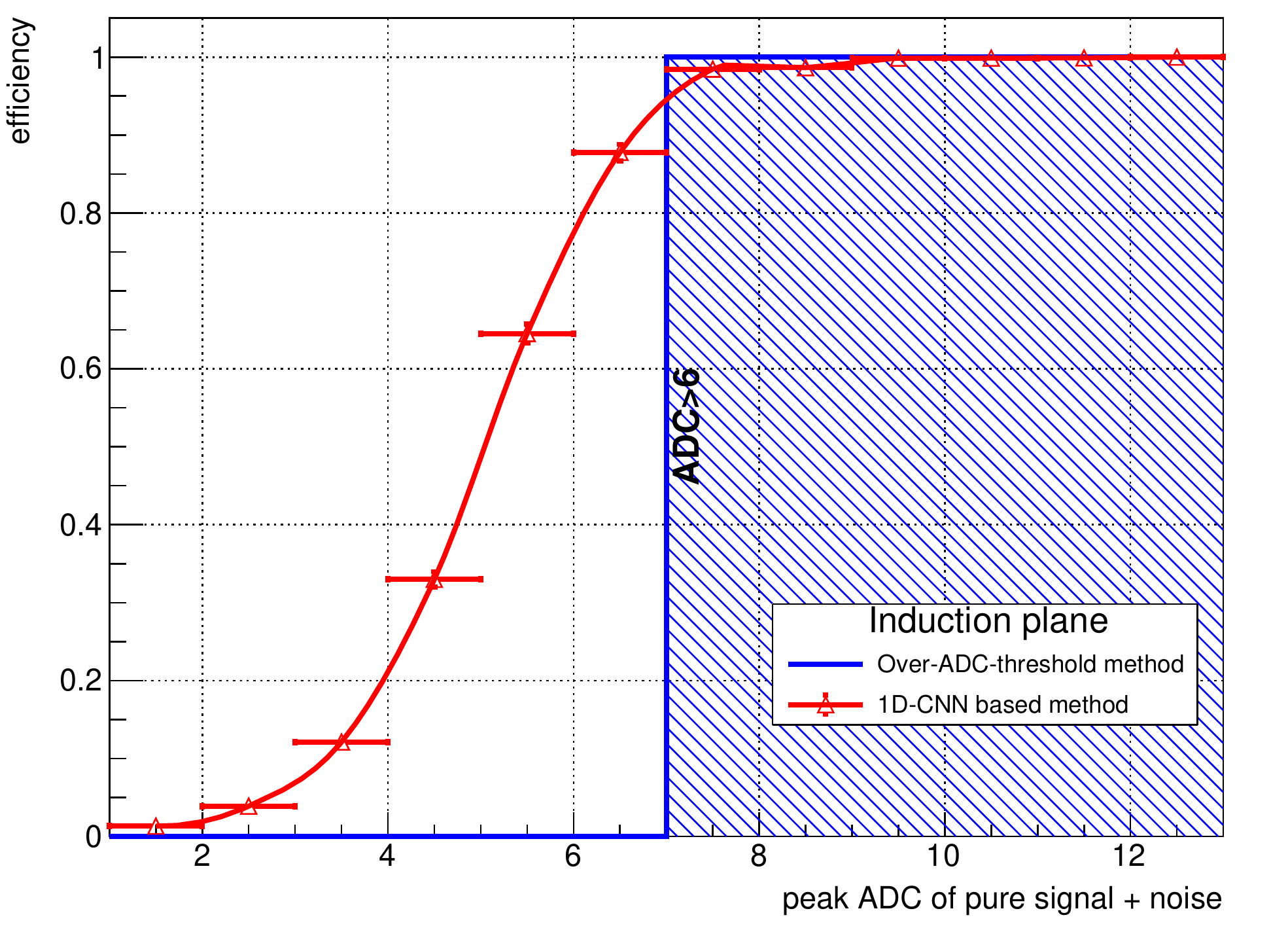} &
    \includegraphics[width=0.49\textwidth]{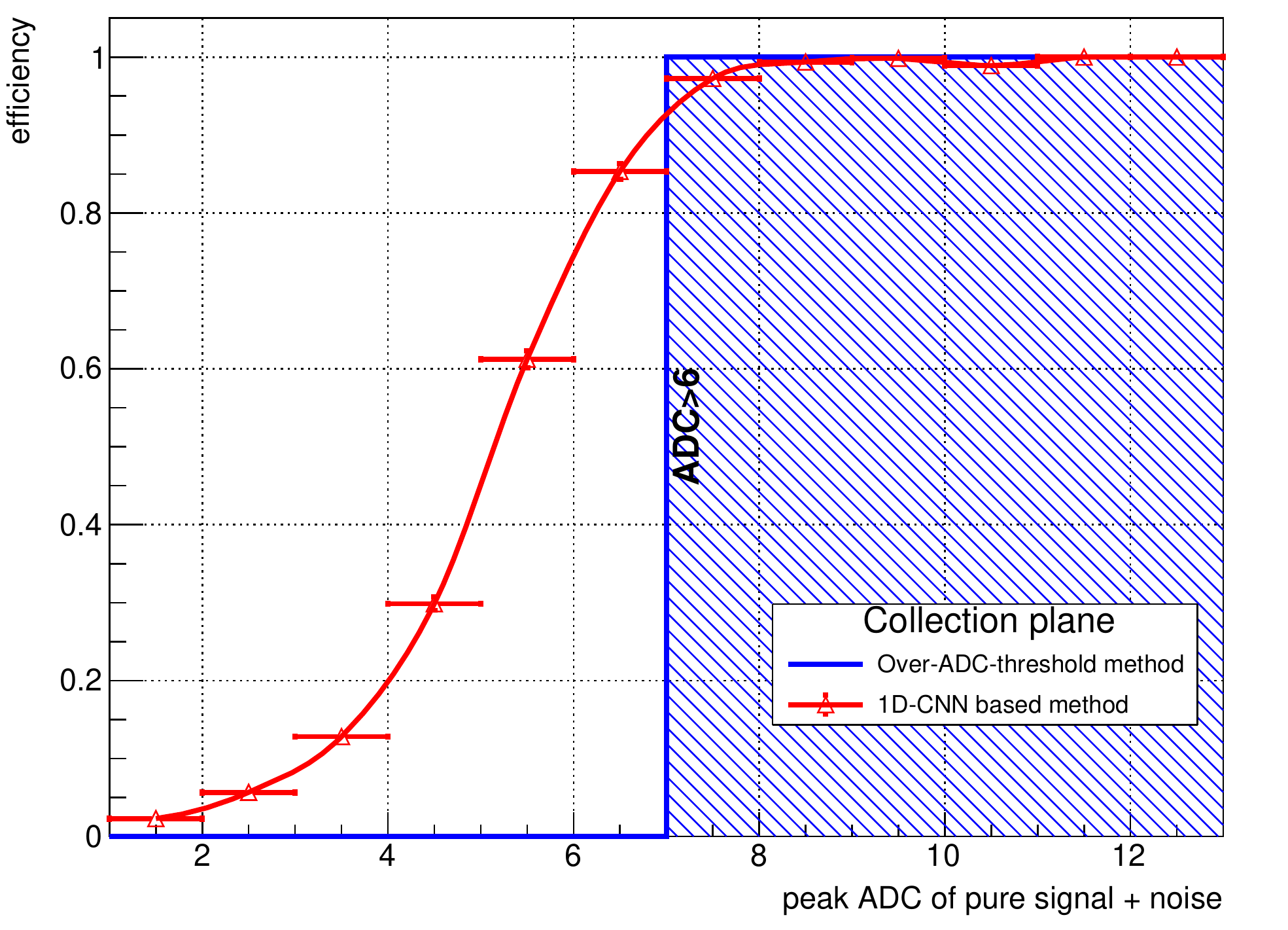} \\
    (a) & (b) \\
  \end{tabular}
\caption{Efficiencies of the 1D-CNN ROI finder for the (a) induction and (b) collection planes as a function of the actual ADC value produced by the detector, which is the sum $ADC_\mathrm{pk}+\mathrm{noise}$.  The blue hatched region represents the efficiency of the \emph{over-ADC-threshold} method.}
\label{fig:roieffvis}
\end{figure*}

The fact that the efficiency plots for the \emph{over-ADC-threshold} method do not exhibit a sharp cutoff at $ADC_\mathrm{pk}=6$ might seem counter-intuitive at first.  However, this is to be expected because $ADC_\mathrm{pk}$ represents the ADC value associated with the pure signal prior to any noise fluctuations.  A signal with $ADC_\mathrm{pk}\le6$ could satisfy the requirement of $\mathrm{ADC}>6$ if it fluctuates above this threshold after the addition of noise.  To demonstrate more clearly how the 1D-CNN is able to detect signals in regions inaccessible to the traditional approach, the 1D-CNN efficiencies are shown as a function of the actual ADC value produced by the detector ($ADC_\mathrm{pk}+\mathrm{noise}$) in Figure~\ref{fig:roieffvis}.  The blue hatched region represents the region above the ADC cut that is accessible to the \emph{over-ADC-threshold} method.

It is not difficult to see from Figure~\ref{fig:roieff} that the CNN-based method outperforms the traditional approach throughout the range from $ADC_\mathrm{pk}=2$ to 8, after which both achieve essentially 100\% efficiency.  The former begins to achieve better than 90\% efficiency at $ADC_\mathrm{pk}=6$, while the latter only reaches this level at around $ADC_\mathrm{pk}=8$.

\section{Conclusion}
In this work, we have successfully demonstrated that deep learning methods can be applied directly to the raw wire waveforms produced by the individual channels in a LArTPC detector in order to discriminate signal waveforms from background.  This was achieved using a 1D-CNN which was implemented in a way that allowed the ROI of the signal to be identified within the full waveform.  The discriminating power of CNNs derives from their ability to learn and detect subtle features that distinguish signal from background such as, but not limited to, shape characteristics.  Because of this, they are not constrained by user defined cuts imposed in traditional threshold based signal ROI finders and can maintain signal sensitivity in energy regions inaccessible to such methods.  The implications are significant for the rich low-energy physics program of future neutrino experiments like DUNE.

A major advantage of our approach over other ML-based ones, is its use of a simple neural network architecture consisting of a mere $\approx$20K trainable parameters compared with the millions found on typical designs.  The low resource utilization makes it feasible for deployment near the detector front end or in edge applications, as part of an online trigger system or for implementing an intelligent zero-suppression filter. Such applications of this 1D-CNN at the very early stages of the DAQ system can help achieve optimal signal efficiency and background rejection that will surely benefit and complement downstream selection algorithms, including more complex ML-based ones.

This paper set out to establish a proof-of-concept and provide a detailed description of the methodology, and it has succeeded in this task.  In subsequent and related papers, this method will be used in actual applications, including those involving real experimental data~\cite{ref:wanweicnnroi}.

\section{Acknowledgements}
We wish to thank Brian Nord and Gabe Perdue of Fermilab for useful discussions on machine learning techniques. This manuscript has been authored by Fermi Research Alliance, LLC under Contract No. DE-AC02-07CH11359 with the U.S. Department of Energy, Office of Science, Office of High Energy Physics. This work has received funding from the European\ Union's Horizon 2020 Research and Innovation Programme under grant agreement number 858199, ``INTENSE".

\bibliography{dlroi_refs}
\end{document}